\documentclass[superscriptaddress,runinaddress,showpacs,showkeys,aps,prd,floatfix,onecolumn]{revtex4}
\usepackage{amssymb}
\usepackage{amsmath}

\usepackage{todonotes}
\usepackage{graphicx,setspace}
\usepackage{dcolumn}
\usepackage{bm}
\usepackage[mathlines]{lineno}
\usepackage[colorlinks=true,citecolor=blue,linkcolor=blue,anchorcolor=green, anchorcolor=green,urlcolor=blue]{hyperref}
\usepackage{palatino}
\usepackage{mathrsfs}
\usepackage{hyperref}
\usepackage{xcolor}
\usepackage{float}
\usepackage{ulem}
\usepackage{lipsum}

\let\originalleft\left
\let\originalright\right
\renewcommand{\left}{\mathopen{}\mathclose\bgroup\originalleft}
\renewcommand{\right}{\aftergroup\egroup\originalright}
\mathcode`\*="8000
{\catcode`\*=\active\gdef*{\mathclose{}\,\mathopen{}}}

\newcommand{\be}{\begin{equation}}
\newcommand{\ee}{\end{equation}}
\newcommand{\bea}{\setlength\arraycolsep{2pt} \begin{eqnarray}}
\newcommand{\eea}{\end{eqnarray}}

\begin{document}

\title{Black holes in the turbulence phase of viscous  rip cosmology}

\author{Iver Brevik}
\affiliation{Department of Energy and Process Engineering, Norwegian University of Science and Technology, N-7491 Trondheim, Norway}

\author{Mubasher Jamil}
\affiliation{Department of Mathematics, School of Natural Sciences
(SNS), National University of Sciences and Technology
    (NUST), H-12, Islamabad, Pakistan}

\date{\today}

\begin{abstract}

We study the phantom fluid in the late universe, thus assuming the equation of state parameter $w$ to be less than $-1$. The fluid is assumed to consist of two components, one laminar component $\rho$ and one turbulent component $\rho_T$, the latter set proportional to $\rho$ as well as to the Hubble parameter, $\rho_T =3\tau H\rho$ with $\tau$ a positive constant associated with the turbulence. The effective energy density is taken to be $\rho_e= \rho + \rho_T$, and the corresponding effective pressure is $p_e=w \rho_e$, with $w$ constant. These basic assumptions lead to a Big Rip universe; the physical quantities diverging during a finite rip time $t_s$.  We then consider the mass accretion of a black hole in such a universe. The most natural assumption of setting the rate $dM/dt$ proportional to $M^2$ times the sum $\rho_e+p_e$, leads to a negative mass accretion, where $M(t)$ goes to zero linearly in $(t_s-t)$ near the singularity. The Hubble parameter diverges as $(t_s-t)^{-1}$, whereas $\rho_e$ and $p_e$ diverge as $(t_s-t)^{-2}$. We also discuss other options and include, for the sake of comparison, some essential properties of mass accretion in the early (inflationary) universe.
\end{abstract}

\maketitle

\section{Introduction}

The observable universe is in the phase of accelerated expansion presumably driven by dark energy carrying negative pressure $p<0$, positive energy density $\rho>0$ and the state parameter $w=p/\rho<0$ \cite{planck}. Among various candidates of dark energy, the phantom energy $(w<-1)$ is the most esoteric one as it violates the weak and null energy conditions. The recent astrophysical data suggests that $w=-1.04^{+0.09}_{-0.10}$ \cite{6}. If phantom energy does exist, than its energy density will increase with time and scale factor will go to infinity in a finite time causing the event known as Big Rip which is a sort of future singularity. Brevik et al proposed that the cosmic fluid will undergo a turbulent phase under extreme cosmic expansion \cite{A}.

There exists at present quite a big number of studies on this kind of future singularity, also with the inclusion of viscosity and even of turbulence. Some references to the Big Rip literature are given in \cite{caldwell03,nojiri03,elizalde04,hao05,brevik05,sola05,brevik11,brevik13,velten13,brevik15}. Another, and milder, variant of the future singularity is called the Little Rip. Then the future singularity is in principle reached, but the universe needs an infinite amount of time to reach the limit $w=-1$.  Some references to the Little Rip can be found in \cite{A,frampton11,frampton11a,brevik11a,bamba12,brevik12, elizalde12}.

The evolution of black holes in a dark energy dominated universe is an active direction of research \cite{F,G,G4,G5}. In particular near the future rip singularity, it is expected that black holes will also go under ripping apart i.e. evaporating much like Hawking radiation culminating in a naked singularity \cite{G1,G2,G3}. This entails the violation of cosmic censorship conjecture itself. We wish to investigate the rate of change of mass of the black hole by taking the effective approach. As it will turn out, the accretion of mass becomes negative in the phantom fluid; the black hole mass $M(t)$ diminishes to zero in a finite time $t_s$, the same time as for the occurrence of the Big Rip itself ($H\rightarrow \infty$).

The turbulent phase, of course, is connected with viscosity. Usually, in a cosmological context one limits oneself to the {\it bulk} viscosity only, because of the assumed spatial isotropy of the cosmic fluid. In the late universe, however, in the phantom era in the vicinity of the future singularity, one must expect that the Reynolds number can be very high and so the {\it shear} viscosity becomes dominant at a local scale. This implies naturally the transition to a turbulent epoch, which we will assume to be isotropic. Typically, in high-Reynolds number isotropic turbulence there exists an inertial subrange in which  the spectral distribution of energy is known, $E(k)=\alpha \epsilon^{2/3}k^{-5/3}$, where $\alpha \approx 1.5$ is the Kolmogorov constant and $\epsilon$ the mean energy dissipation per unit time and unit mass. When $k$ becomes as large as the inverse Kolmogorov length, $k \rightarrow (\epsilon/\nu^3)^{1/4}$ with $\nu$ the kinematic viscosity, the fluid enters into the dissipative region. We will formulate simple assumptions for the turbulent phase below.

When considering the accretion rate $dM/dt$
 for a black hole in the phantom universe, it is in our opinion natural to put this rate proportional to the sum of the effective energy  density and effective pressure.   Based upon this fundamental assumption, we find that the mass  decreases,  as mentioned, going to the limit $M=0$ in a finite time. Detailed calculations on this case are presented in Section II. We also consider briefly other options, for the strength of the turbulent fluid component, and for the form of $dM/dt$.

For the sake of comparison, we make in Section III some comments also on black holes in the early (inflationary) universe. Also in that case it has been found that negative mass accretion can  occur, although then due to the interaction with a scalar field.

\section{ Modeling}

Before embarking on the case of turbulent fluids, let us make some comments on the use of laminar viscous fluid models in cosmology. The introduction of a bulk viscosity was made a long time ago, a classic reference being the 1971 paper of Weinberg \cite{weinberg71}.  The topic has later attracted considerable interest; a recent review can be found, for instance, in Ref.~\cite{brevik17}.  Assuming for simplicity a one-component  fluid, with the equation of state of the conventional form
\begin{equation}
p=w \rho,
\end{equation}
with $w$ a constant. If the fluid is in the phantom region, $w$ is less than $-1$.

 It has turned out that the following phenomenological formula for the density-dependent bulk viscosity,
\begin{equation}
\zeta = \zeta_0\left( \frac{\rho}{\rho_0}\right)^{\lambda},
\end{equation}
with a fixed $\lambda$, is useful for comparison with experiments. Here we assume for definiteness that subscript zero refers to the present time, $t=0$. In view of the experiments  one can conclude  shown that $\lambda =1/2$ is most likely an optimal choice for the exponent \cite{normann16,normann17}. In that case it  moreover turns out  that the analytic expression for the redshift-dependent Hubble parameter becomes quite simple, \cite{normann16}
\begin{equation}
H(z)=H_0\sqrt{\Omega(z)}(1+z)^{-B/H_0}, \label{H}
\end{equation}
where
\begin{equation}
\Omega(z)=\Omega_0\, (1+z)^{3(1+w)}, \quad \Omega_0=\frac{\rho_0}{\rho_c}, \quad B= \frac{3}{2}\zeta_0.
\end{equation}
Here $\rho_c=3H_0^2$ is the critical density (we use units for which $8\pi G=1$).

 We will consider the transition to a state where part of the effective energy density $\rho_{\rm eff}$ exists in the form of an isotropic turbulent fluid. Thus $\rho_{\rm eff}=\rho+\rho_T$. We let the onset of turbulence occur at the time $t=0$. The influence from turbulence may be divided roughly into two classes: moderate, or strong, turbulence.
\bigskip

\noindent {\bf Moderate turbulence.} Assume, as in Ref.~\cite{A}, that $\rho_T$ is proportional to $\rho$, as well as to the scalar expansion $\theta$. Calling the proportionality constant $\tau ~ (>0)$, we thus have $\rho_T=\rho\times \tau \theta=3\tau \rho H$, and so
\begin{equation}
\rho_e=\rho(1+3\tau H). \label{1}
\end{equation}
Similarly we write he effective pressure as $p_{\rm eff}=p+p_T$, where $p=w\rho$ as above for the laminar part, and $p_T=w_T\rho_T$ with $w_T$ another constant, in principle different from $w$.  For simplicity we will however set the two thermodynamic constants equal, $w_T=w$, so that the effective pressure takes the  form
\begin{equation}
p_e=w\rho (1+3\tau H), \label{1a}
\end{equation}
analogous to the energy equation (\ref{1}).

The governing equations are \cite{A}:
\begin{eqnarray}
\dot H&=&-\frac{1}{2}(\rho+p),  \\
H^2&=&\frac{1}{3}\rho(1+3\tau H),\\
2\dot H+3H^2&=&-w \rho(1+3\tau  H)
\end{eqnarray}
This system of equations is easy to solve.
We define $\gamma$ as  $\gamma=1+w$, and insert a generalized ansatz for the turbulent energy density, $\rho_T=3\tau \rho H^n$, with $n$ an integer. Then we obtain the equation
\begin{equation}
(1+3\tau H^n)(\dot{H}+\frac{3}{2}\gamma H^2)=0,
\end{equation}
which has the solutions
\begin{equation}
H_1(t)=\Big(\frac{-1}{3\tau}\Big)^{-1/n},~~~ H_2(t)=\frac{2}{3\gamma t-2 C_1},
\end{equation}
where $C_1$ is a constant of integration. As the Hubble parameter cannot be negative, the solution  $H_1$ is non-physical unless  $\tau <0$. That de Sitter-like solution is physically unreasonable,  however, because it would imply  the turbulent part of the energy to be  negative. We therefore focus on the solution $H_2(t)$ in the following, calling it $H(t)$. The corresponding scale factor (for $n=1$) becomes
  \begin{equation}
a(t)=a_0 Z^{2/3\gamma}, ~~~Z=1+\frac{3}{2}\gamma H_0 t, \label{12}
\end{equation}
implying that
\begin{equation}
H=\frac{H_0}{Z}. \label{12a}
\end{equation}
 Consequently, the sum of effective energy density and pressure is
\begin{equation}
\rho_e+p_e=\frac{3H_0^2 \gamma}{Z^2} \label{12b}
\end{equation}
To track the evolution of black hole mass in turbulent phase, we has to  establish an   accretion equation. The form of such an equation  is not evident beforehand, in view  of the insufficient knowledge about the phantom universe. A mild restriction, which we shall adopt henceforth, is  to take the accretion rate $dM/dt$ to be proportional to $M^2$ through some time-dependent function that we shall call $f(t)$,
\begin{equation}
\frac{dM}{dt}=M^2 f(t).
\end{equation}
The proportionality to $M^2$ comes from integration of the radial flux over the whole area of the black hole; cf. Refs. \cite{F,babichev04,azreg-ainou18}.

We consider the mass accretion to be quasi-static.
There are two options for the choice of $f(t)$ that appear natural: we make take $f(t)$  to be proportional to the sum of the effective energy density and pressure; or we may take it to be proportional to the effective density only. We will now outline what are the consequences  of these options.

\subsection{Assume $f(t)$ proportional to $(\rho_e+p_e)$}

We will first assume that   $\rho_e$ and $p_e$  act on the same level, and write
\begin{equation}
f(t)=4\pi A (\rho_e+p_e)  = 4\pi \gamma A \rho_e,  \label{X}
\end{equation}
where  $A$ is an unspecified constant. This assumption is of course simple, but it seems to conform with the same role played by energy density and pressure in general relativity. (Compare, for instance, with the energy conservation equation $\dot{\rho}+3H(\rho+p)=0$ in standard notation.)
Thus
\begin{equation}
\frac{dM}{dt}=\frac{12\pi A M^2H_0^2 \gamma}{Z^2}, \label{Y}
\end{equation}
which means that the mass decreases in the phantom domain ($\gamma < 0$. The solution is
\begin{equation}
M(t)=\frac{M_0Z}{8\pi A H_0 M_0(1-Z)+Z} \label{16}
\end{equation}
(notice that $\gamma$ is here present only in the $Z$ term). The mass decreases to zero in a finite time $t_s$, where
\begin{equation}
t_s=\frac{2}{3|\gamma|H_0}.\label{Z1}
\end{equation}
The turbulence parameter $\tau$ does not occur in these expressions; this is a consequence of the ansatz (\ref{X}) which involves the effective quantities $\rho_e$ and $p_e$ only.

Thus, at $t=t_s$, both
\begin{equation}
H=\frac{2}{3|\gamma|(t_s-t)},
\end{equation}
\begin{equation}
\rho_e=3H^2= \frac{4}{3\gamma^2(t_s-t)^2},
\end{equation}
and $p_e=w \rho_e$, diverge. These are characteristic properties for the Big Rip singularity. Notice the contrast with   the Little Rip singularity in which an  infinite time is needed to reach infinity (for a classification of future singularities, see Ref.~\cite{nojiri05}).

What is the time dependence of the laminar energy density component $\rho$? From the definition equation (\ref{1}) we obtain
\begin{equation}
\rho=
\frac{4}{3|\gamma|(t_s-t)}\frac{1}{2\tau+|\gamma|(t_s-t)},     \label{Z4}
\end{equation}
which shows that the singularity of $\rho$ is only of order of $1/(t_s-t)$. It is thus weaker than that of $\rho_e$.

For the mass itself we get
\begin{equation}
M(t) \rightarrow \frac{3}{16}\frac{|\gamma|}{\pi A}(t_s-t), \quad  t \rightarrow t_s,
\end{equation}
showing that the behavior is just linear in $(t_s-t)$ near the Big Rip. This behavior is influenced solely by the phantom fluid and is not related to any Hawking radiation.

Let us make an estimate of the magnitude of the rip time, taking as an example
\begin{equation}
\gamma = -0.05.
\end{equation}
Then, with $H_0=67.74$ km~s$^{-1}$~Mpc$^{-1}$ = 2.20$\times 10^{-18}$ s$^{-1}$, we get
\begin{equation}
t_s \approx 190~\rm{Gyr},
\end{equation}
which is a typical magnitude for the rip time.

\bigskip

{\bf Strong turbulence.}  Instead of assuming the turbulent component in the fluid to be moderate, as represented by Eqs.~(\ref{1}) and (\ref{1a}), we will now explore the consequences of a more strong assumption, namely to let  $\rho_e$ and $p_e$  be attributed to the {\it Hubble parameter} only, without any reference to the laminar fluid component $\rho$ at all. That means physically that the turbulence is taken to be the overwhelming factor in the late phantom fluid. Without changing the symbol $\tau$ in the formalism, we may thus write the effective quantities as
\begin{equation}
\rho_e=3\tau H, \quad p_e= 3w\tau H.  \label{W}
\end{equation}
In this case the Friedmann equations yield
\begin{equation}
H=\tau,
\end{equation}
\begin{equation}
2\dot{H}+3\gamma \tau^2=0.
\end{equation}
This implies that  $\gamma=0~(w=-1).$ The occurrence of the phantom-divide equation of state thus follows from the assumption (\ref{W}) about turbulence, and is not a condition imposed initially by hand. The universe will  develop in a de Sitter phase where the Hubble parameter is completely determined by the  parameter $\tau$.  The scale factor develops as
\begin{equation}
a(t)=a_0e^{\tau t}. \label{ZZ}
\end{equation}
As $\rho_e+p_e=0$ in this case, the mass of the black hole will stay constant, $M=M_0$. The effective energy pressure and pressure become in this case $\rho_e=3\tau^2,~ p_e=3w\tau^2$. The fluid is completely dominated by the turbulence parameter $\tau$.

\subsection{Assume $f(t)$ proportional to $\rho_e$}

As mentioned above, this is also an option that is worth investigating. Instead of Eq.~(\ref{Y}) we now get
\begin{equation}
\frac{dM}{dt}=\frac{12\pi A M^2H_0^2 }{Z^2}. \label{Y1}
\end{equation}
Starting again from the value $M=M_0$ at $t=0$, it is convenient to write the solution as
\begin{equation}
\frac{M_0}{M}=1-\frac{12\pi AH_0^2M_0t}{1-t/t_s}.
\end{equation}
with $t_s$ still the rip time as given by Eq.~(\ref{Z1}).

The behavior of this expression is complicated. The mass increases with respect to $t$, and reaches the limit $M \rightarrow \infty$ at a finite time $t=t_g$, where
\begin{equation}
t_g= \frac{t_s}{1+12\pi AH_0^2M_0t_s}. \label{W2}
\end{equation}
Thus $t_g<t_s$, so the limit $M\rightarrow \infty$ occurs before the rip time. Where does this accretion of energy come from? The only plausible explanation is that it is extracted  from the phantom fluid itself. However, the kind of behavior shown by Eq.~(\ref{W2}) is after all strange, and we conclude that the basic assumption (\ref{Y1}) is less physical than  that of the previous subsection.

\section{Comparison with the early universe}

For the sake of comparison, let us briefly review some characteristics of the theory of black hole evaporation in the {\it early} universe. In that case, it is known that
\begin{equation}
\frac{dM}{dt}=\left(\frac{3}{4}\right)^3 \frac{M^2}{\pi}\rho_{\rm rad},
\end{equation}
(recall $8\pi G=1$), where $\rho_{\rm rad}$ is the radiation energy density far outside a Schwarzschild black hole (cf., for instance, Refs.~\cite{shapiro83,E}). This result follows in turn from the capture cross section for photons coming from infinity,  $\sigma_{\rm phot}=(3/4)^3 M^2/\pi$. This form is thus different from the basic form  (\ref{X}) that we assumed above for the late universe.

Continuing with the early universe, at inflationary stages
\begin{equation}
\rho_{\rm rad}=3H^2=\frac{3}{4t^2},
\end{equation}
since $H=1/(2t)$.
Thus
\begin{equation}
\frac{dM}{dt}=\left(\frac{3}{4}\right)^4\frac{M^2}{\pi}\frac{1}{t^2}, \label{Z}
\end{equation}
with here $t=0$ referring to the Big Bang.

We  may integrate Eq.~(\ref{Z}), starting from an initial time $t=t_0$, to get
\begin{equation}
M(t)= \frac{M_0}{1+\left( \frac{3}{4}\right)^4\frac{M_0}{\pi}\left( \frac{1}{t}-\frac{1}{t_0}\right)} .
\end{equation}
Thus $M=M_0$ at $t=t_0$, while $M$ increases with time and approaches a finite value when $t\rightarrow \infty$.

It is moreover worthwhile to notice that the negative accretion of mass is found also in the early universe theory. Thus Rodrigues and Saa \cite{rodrigues09} considered the evolution of the mass of  Schwarzschild black holes in the presence of a nonminimally coupled scalar field, and found that any black hole with initial mass $M_0$ will disappear due to the accretion of the {\it scalar} field. This was found to hold true even in the absence of Hawking radiation. The point we wish to emphasize here is that negative accretion can place both in the early, and in the late,  universe, although via different mechanisms.  In the case considered in Ref.~\cite{rodrigues09}, the disappearance of the black hole would require an infinite amount of time.

\section{Summary}

The concept of viscosity has attracted considerable attention in modern cosmology. For a homogeneous fluid with energy density $\rho$, the bulk viscosity $\zeta \propto \rho^{\lambda}$ with the constant $\lambda$ preferably equal to 1/2, turns out to  a very useful ansatz, as far as one can judge on the basis of experimental information (cf., for instance, \cite{normann16,normann17}). It corresponds to a Hubble parameter $H(z)$ that  is given by Eq.~(\ref{H}) above, as function of the redshift.

The main objective of the present work was however to abandon the assumption about a homogeneous cosmic fluid and allow for an isotropic turbulence component.  We  focused on the late universe, assuming that the equation of state parameter $w$ is constant and less than $-1$. This means a phantom fluid. A natural approach is to write the effective energy density $\rho_e$ as a sum of two components, one laminar component $\rho$ and one turbulent component $\rho_T$, where the latter is taken to be   proportional to the Hubble parameter, $\rho_T =3\tau H\rho$ with $\tau$ a positive  constant induced by the turbulence. The expression for $\rho_e$ is  given  in Eq.~(\ref{1}). The corresponding effective pressure is written as $p_e=w\rho_e$. This is the same model as we introduced earlier, in Ref.~\cite{A}. It corresponds to a moderate degree of turbulence.
Based upon these assumptions, we found by use of the Friedmann equations that the scale factor $a(t)$ diverges at a finite rip time $t_s$;  cf.  Eq.~(\ref{12}). This is characteristic for the Big Rip singularity.

We  went  on to  examine the accretion of mass of a black hole in the phantom universe, starting from an initial mass $M=M_0$ at an arbitrary time $t=0$. Writing the rate as $dM/dt =M^2 f(t)$, we adopted first the natural choice of setting  $f(t) \propto (\rho_e+p_e)$. That means, we assumed   the effective energy density and effective pressure to act on the same footing. The result, shown in Section II.A, was that the mass accretion is negative; $M(t) \rightarrow 0$ linearly in $(t_s-t)$ near the singularity. Otherwise, for the effective energy density and pressure, the Big Rip characteristics were recovered. The laminar energy density component $\rho$, however, was found to obey a milder singularity than $\rho_e$; cf. Eq.~(\ref{Z4}).

In Section II.A we also investigated the extreme case of strong turbulence, putting $\rho_e$ and $p_e$ simply proportional to $H$. The result was quite a different evolution of the universe, being of de Sitter type with a Hubble parameter simply equal to $H$.

Section II.B  dealt with the ansatz $f(t) \propto \rho_e$. This case was found to yield positive mass accretion, but the solution for $M(t)$ showed a strange behavior making us conclude that this option is of less physical interest.

In the final Section III, we made some simple comparisons with the evolution of a black hole in the {\it early}  universe.  At the inflationary stage, $M(t)$ is typically found to increase with time until a stationary state is approached. However, also in the early universe, negative mass accretion has been found to occur \cite{rodrigues09}.

\end{document}